# Recorded Accounts of Meteoritic Events in the Oral Traditions of Indigenous Australians

Duane W. Hamacher

Nura Gili Indigenous Programs, University of New South Wales, Sydney, NSW, 2052, Australia
Email: d.hamacher@unsw.edu.au

**Abstract**

Descriptions of natural events, such as fireballs and meteorite impacts, are found within Indigenous Australian oral traditions. Studies of oral traditions demonstrate that they extend beyond the realm of myth and legend; they contain structured knowledge about the natural world (science) as well as historic accounts of natural events and geo-hazards. These traditions could lead to the discovery of meteorites and impact sites previously unknown to Western science. In addition to benefiting the scientific study of meteoritics, this study can help social scientists better understand the nature and longevity of oral traditions and further support the growing body of evidence that oral traditions contain historical accounts of natural events. In a previous study led by the author in 2009, no meteorite-related oral traditions were identified that led to the discovery of meteorites and/or impact craters. This paper challenges those initial findings.

**Warning to Aboriginal Readers:** This paper contains the names of Aboriginal people who have passed away.

## INTRODUCTION

The study of meteoritic phenomena in Indigenous oral traditions has been a topic of research interest for several years, particularly in Australia (see Bevan and Bindon 1996; Goldsmith 2000, Hamacher 2011, Hamacher and Goldsmith 2013, Hamacher and Norris 2009, 2010, 2011a). Research in the growing discipline of *geomythology* - a close cousin of ethnoastronomy - involves investigating oral traditions for descriptions of past geologic events (Vitaliano 1973) that may provide insight into both the culture that observed and recorded them, and for information about the event that might contribute to our understanding of geological phenomena. The study of geomythology provides important insights into how natural events are understood and incorporated into oral traditions, as well as providing direct methods for dating oral traditions.

Hamacher and Norris (2009) published a treatise on meteorite falls and impact events in Indigenous Australian oral traditions. One of the main goals of the research was to determine if meteoritic events were recorded in oral tradition and whether these traditions would lead to the (re)discovery of meteorites and impact structures previously unknown to Western science. In the 2009 study, no oral traditions accounts were found that led to the discovery of meteorites or craters. Additionally, no oral traditions were identified from Queensland or Victoria except for an account of Aboriginal people interacting with one of the Cranbourne meteorites near Melbourne and a possible account of a meteorite fall in the Torres Strait. No evidence at the time connected any of the Indigenous traditions of impact events to known impact sites.

Research since 2009 reveals more Indigenous accounts of meteoritic events from across Australia, particularly Queensland and Victoria. Analysis reveals that meteorites and craters recorded in oral tradition were later verified by Western science.

The purpose of this study – like the 2009 study – is not to validate or legitimize Indigenous oral traditions, as they do not need to be validated or legitimized. Instead, this study tests three hypotheses:

1. Indigenous oral traditions contain historical accounts of meteoritic events;
2. Indigenous oral traditions can lead to the identification of meteorite falls and impact craters previously unknown to Western Science; and
3. Scientific studies of these falls and/or craters can help us understand the nature and longevity of Indigenous oral traditions.

| **Name** | **State** | **Latitude** | **Longitude** | **Type** | **Found** |
|---|---|---|---|---|---|
| Benyeo | VIC | −38.283 | 141.616 | | |
| Carnarvon | WA | −24.884 | 113.657 | | |
| Helidon Springs | QLD | −27.549 | 152.099 | | |
| Henbury | NT | −24.572 | 133.148 | IIIA | 1931 |
| Lake Argyle | WA | −16.361 | 128.748 | | |
| Lake Macquarie | NSW | −32.995 | 151.600 | | |
| Lilydale | VIC | −37.769 | 145.341 | | |
| Monte Colina | SA | −29.400 | 139.983 | L3 | 1963 |
| Munpeowie | SA | −29.583 | 139.900 | IC | 1909 |
| Narraburra | NSW | −34.367 | 147.878 | IIIB | 1855 |
| Saibai | QLD | −09.381 | 142.615 | | |

**Table 1**: *Places described in the results, including the name, location, meteorite and year found (if known). Meteorite locations, types, and years found from Bevan (1992).*

**RESULTS**

Using the methodologies of the Hamacher and Norris (2009) study, this paper reveals eight oral traditions that describe meteoritic events and three minor accounts that describe meteoric phenomena or the presence of a meteorite by Indigenous people. Each of these accounts are recorded in the literature as having special significance to Indigenous Australians, providing both a location and description of the event. These are then analysed to determine if they support of reject the hypotheses described above. The locations of each account (Table 1) are used to determine if any meteorites were recovered from the region using Bevan (1992), Gibbons (1977), Grady (2000), and (Meteoritical Society 2013).

Places mentioned in the text are shown on a map of Australia in Figure 4, with those mentioned in the Torres Strait shown in Figure 1 (in order of their appearance in the paper).

***Henbury Meteorites Conservation Reserve, Northern Territory***

In the 2009 study by the author (Hamacher and Norris 2009:66-67), there was little evidence that the formation of the Henbury crater field, which occurred < 4,700 years BP, was recorded in oral tradition. The only suggestion was in the name (*chindu china waru chingi yabu*, roughly translating to "sun walk fire devil rock"), which vaguely suggested a living memory of the event. Recent research by Hamacher and Goldsmith (2013:299-303) uncovered additional records relating to Aboriginal views of the site and these records clearly indicate a living memory of the impact recorded in oral traditions.

The name "*chindu china waru chingi yabu*" is of the Luritja language. Historical documents give more information about Aboriginal views of the site. When James M. Mitchell visited the site in 1921, he took an Aboriginal guide. His interest was piqued when his guide refused to go near them, saying that it was a place where a fire "debil-debil" [devil] came out of the sky and killed everything in the vicinity. He visited the craters again in 1934 and took another Aboriginal guide with him. The guide said Aboriginal people would not camp within two miles of the craters or even venture within half a mile of them, describing them as a place where the fire-devil lived. He claimed they did not collect water that filled some of the craters, fearing the fire devil would fill them with a piece of iron. The guide said his grandfather saw the fire devil and it came from the sun. Aboriginal groups to the north of Henbury (including the Kaitish and Warramunga) hold traditions that meteors are fiery "debil-debils" that hurtle from the skies to feast upon the entrails of the recently deceased (Hill 1937).

In March 1932, an unnamed resident of the area undertook independent research and spoke to local Aboriginal elders. According to the elders, all young Aboriginal people were forbidden from going near the craters. The elders described them as the place where "a fiery devil ran down from the sun and made his home in the Earth. The devil will kill and eat any bad blackfellows," (see Hamacher and Goldsmith 2013:300-301).

These varied (but similar) accounts, recorded over a period of 10 years, seem to confirm the presence of an oral tradition describing the impact formation of the craters. There is the question of cultural contamination, i.e. colonial scientific interest in the site influencing Aboriginal traditions, but there is no evidence that this occurred. The consistency of the traditions over the 10-year time period in which they were recorded supports the hypothesis that the traditions were pre-colonial, but we cannot know either way for certain. By 1945, colonial interest in the craters led to Aboriginal people collecting and selling pieces of the "star that fell to Earth" (Vox 1945), indicating outside influence by this point.

These accounts stand in contrast to claims in the literature that no oral traditions of the Henbury impact can be found and that the crater field was of no interest to local Aboriginal people (Alderman 1932, Anonymous 1934). The current evidence indicates that Aboriginal people witnessed the event, recorded the incident in oral traditions, and those traditions remained intact through the 1930s (and possibly later).

*Narraburra, New South Wales*

The Narraburra meteorite (aka Yeo Yeo), is a 32.2 kg octahedrite found near Stockinbingal, NSW in 1855 (Hodge-Smith 1939:23). According to Henry Chamberlain Russell, who served as the Government Astronomer from 1870-1905, the ox-skull-shaped meteorite was found lying on hard, stony ground by a Mr. O'Brien (first-name not given), then passed on to Patrick Herald, then to Russell in 1890 (Russell 1890:82). The word *Narraburra* is from a Wiradjuri Aboriginal dialect meaning "rough country" (Thorpe and McCarthy 1958:18). Sometime between acquiring the meteorite in 1890 and his death in 1907, Russell recounted the discovery of the meteorite to Gale (1924:4). An abridged version is as follows:

"*Some time back* […] *this aerolite, according to Aboriginal lore, descended on earth and half buried itself there at the head of Bland Creek, which has its principle source in the vicinity of Stockinbingal. It is a huge block of stone* […]. *The blacks were terribly afraid of it, believing it to be possessed of supernatural powers. So they called the strange visitor "Yeo Yeo", a synonym for their pidgin English "Debbil-Debbil"* [Devil-Devil]. *And Yeo Yeo Creek it remains today.*"

Neither Russell nor any other authors mention the Aboriginal story or discovery of the meteorite in their formal publications. If the fall was witnessed by Aboriginal people and handed down over time through oral tradition, how long ago did the fall occur? Liversidge (1903) analysed the meteorite but gave no indication of its age on Earth. Additional analysis of the meteorite revealed that large, hemispherical, undercut cavities on the stone were believed to be corrosion pits that developed during its long exposure to the Earth's atmosphere. Using cosmic ray exposure, Chang and Wanke (1969) estimated the duration of this exposure to be between 150,000 and 340,000 years (see also Buchwald, 1975:876).

According to Russell (via Gale), the discovery of the Narraburra (Yeo Yeo) meteorite was based on Aboriginal oral tradition, and the tradition claims that people witnessed its fall. While it is fairly certain that Aboriginal people would have witnessed meteorite falls and developed traditional stories these events, this is one of the few recorded examples that demonstrates this. It also highlights that the name of Yeo Yeo is based on the meteorite. Doubt is cast on the validity of the supposedly observed fall in light of the discrepancy between the duration of time humans have inhabited Australia and the time that Chang and Wanke (1969:401) estimate the meteorite was exposed to the atmosphere (a gap of 100,000 to 240,000 years). No research about the meteorite's exposure age has been published since 1969, but modern advances in dating techniques (e.g. Dunai 2010) might reveal a more accurate exposure age.

*Strzelecki Regional Reserve, South Australia*

Two reports of meteorites were identified from the sandhills near the sand dune country of the Strzelecki Regional Reserve in northeastern South Australia. Neither report is an oral tradition, but rather an account of Aboriginal people telling non-Indigenous Australians about the locations of supposed meteorites.

The first is a description of a tiny ~2.5 m "crater" in the "Monte Collins" (aka Monte Colina/Collina) sandhills that an Aboriginal man said was formed by a meteorite (Gill 1926). Gill reports that four meteorite fragments were recovered nearby. The Grady (2000:132) catalogue reveals that a small 116.8 g meteorite was identified from Monte Colina in 1963. The nearby 2.8 kg Accalana and 31.6 kg Carraweena meteorites, which are identical and part of the same fall (Heymann 1965) are probably also identical to the Monte Collina meteorite, as they are all part of the rare L3 type (Mason 1974:177). The Artracoona meteorite was found in 1914, 10 km west of the Carraweena and Accalana meteorites, but is distinct from them (*ibid*:79). Therefore, it is plausible that the Monte Colina meteorite is the one described by the Aboriginal man. If it was accompanied by a small crater, it suggests the fall is fairly recent, as such a small structure would have eroded away or filled with debris otherwise. An age of the fall is not given.

The second report is from H.J.L. (1926), who says that Aboriginal people told of a large meteorite in the sandhills west of Monte Barcoola waterhole. The location of Monte Barcoola could not be found, but is 245 km northeast of Hergott Springs, SA, also near Strzelecki. The same Aboriginal people who told of the Monte Barcoola meteorite claimed to have seen the nearby Murnpeowie meteorite "fall to the southwest." The 1,143 kg Murnpeowie meteorite lay 70 m from an elliptical hole in the ground (along the major axis) that is 5.0 m long, 3.7 m wide, and 1.2 m deep. It was reported to have first identified by an unnamed Aboriginal man in 1909 (Anonymous 1910). It was believed that the meteorite came from the west, struck the ground at a low angle, and then ricocheted off to its present location. It is not known when the fall occurred, but analysis (Smith 1910, Spencer 1934) suggested that the impact had not occurred before the erection of a nearby fence just five years earlier. The highly preserved state of the alleged crater supports a young age.

### Benyeo Homestead, Victoria

In 1888, 40-tonnes of ironstone were found near Bringalbert, Victoria. An Aboriginal man named Bobby Fry stated that his father told him the stones "fell from the sky," (Grassie 1888):

*"There is a sand dune on the Benyeo side of Bringalbert, which, contains about forty tons of iron stone - the only iron stone in that quarter above, beneath, or around. Bobby Fry, the aboriginal […] asserts that he heard his old father say that those stones fell one day from the sky and it is possible that they did so. A squatter at Benyeo is having them built into a wing to his castle, and he will be able to boast soon that one of his wings was once a comet."*

Benyeo homestead is northwest of Apsley, Victoria. It was first built in 1863 from local ironstone for the "pioneer-settler" (squatter) Hugh Lawrence McLeod (Victorian Heritage Database, 1974). The homestead possessed half octagonal windows, giving it a bit of a castle-like appearance. It was extended in 1882 – presumably not the last extension as mentioned in the account. It is not known if the extension used meteoritic iron and nothing more is recorded of the ironstone or the meteorite. No meteorites are catalogued from this area (the nearest catalogued meteorite is from Dimboola, Victoria, which was found in 1944). It should be noted that Grassie also recorded a meteorite fall near Bringalbert that "buried itself in the sand" (Grassie 1898), but nothing is recorded in meteorite databases.

*Saibai Island, Queensland*

In the Torres Strait, some landscape features were associated with objects falling from the sky in Islander oral traditions. According to Barham et al. (2004:23), these features include a boulder site called *Daparau Kula* (meaning "the stone that fell from the sky") and rock art sites are associated with unusual landforms that also are attributed to stars falling from the sky, such as those at Keriri (Hammond Island) and the kod site on Pulu Islet near Mabuiag Island (Barham et al. 2004:55). These features are terrestrial in nature but are attributed to cosmic origins. It is possible, however, that some may relate to actual meteorites and meteorite falls.

An oral tradition from Saibai Island describes the fall of a large stone from the sky that remains embedded in the ground (Hamlyn-Harris 1913). Saibai is 4 km from the southern coast of Papua New Guinea. It consists predominantly of fairly flat mangrove swamp (the highest point on the island is 1.7 m above sea level). Near the turn of the 20$^{th}$ century a unique 203 kg stone was identified by a colonist from local oral traditions. According to elder Saibai men, the stone "fell from the heavens," striking the ground near a sitting man where the coastal village now stands. After the near miss, the man rose and fled. Oral traditions claim a second stone fell on Dauan Island (7 km west of Saibai) and killed a number of people. A similar account is described from Pulu islet, near Mabuiag Island (78 km southwest of Saibai) (Haddon 1904:22; see Figure 1).

According to Charles Neibel, the Government Teacher on Saibai Island (*ibid*:5-7):

*"Moigi, a man of about sixty years of age, says that when he was a boy his father Kubid told him the story, which he had heard from his father, Ausi, that the stone in question had fallen from the sky and did not belong to this world. Ausi (the grandfather of Moigi) had not seen it fall — it did not fall during his lifetime, but he had the story as it had been handed down from father to son by his (Ausi's) forefathers. The story being already traditional during the childhood of the grandfather of one of our oldest men, points to the fact that the stone is more than a century old; perhaps considerably more. The stone was allowed to lie where it fell, and, during the childhood of those who are now old men, parents used to forbid their children from touching it, for fear that if they touched it more stones would fall. When the first missionaries came they said their God was the only god and that the stone could not hurt them, and suggested burning it. Then five men — Gari, Dagi, Aina, Janaur, and Kinaur — put fire round the stone, and managed to chip off the outer shell for stone clubs, but could make no impression on the inner portion. By this means they reduced the diameter of the stone by about six or eight inches. After that the stone lost its sanctity and children used to play freely round it and climb on to it."*

The stone was rolled from its *in situ* position to assist in the reclamation of the swamp area in which it was found. Sir William MacGregor, the Governor of Queensland from 2 December 1909 to 16 July 1914, sent the stone to the Queensland Museum for examination. The examination, conducted by John Brownlie Henderson (the Queensland Government Analyst), concluded that the stone was not of meteoritic origin. A subsequent report by Charles Anderson at the Australian Museum in Sydney stated that the term "fell from heaven" indicated Christian

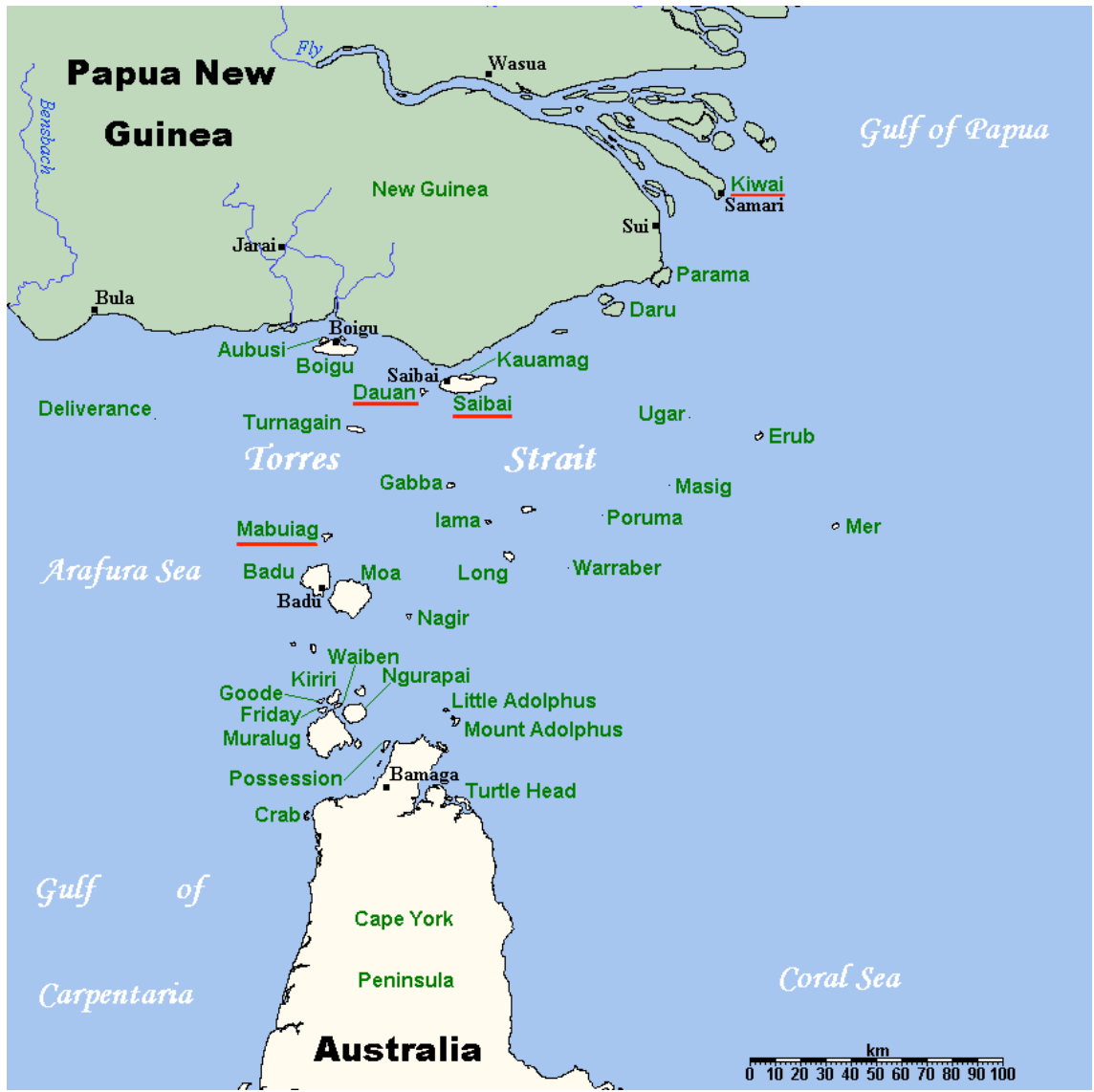

*Figure 1:* Places described in the Torres Strait, underlined in red. Image Kelisi (Wikipedia).

teaching rather than traditional knowledge (Saibai Islanders were converted to Christianity after the arrival of the London Missionary Society in 1871) and concluded that (*ibid*:6):

*"...unless its fall was actually observed by reliable witnesses, I am afraid that the meteoritic origin of this specimen would not be accepted on the evidence of legendary reports. It would be unsafe to say that a body with the characteristics of andesite might not reach the earth from space, but possibilities are not probabilities, and such a substance would have to furnish unexceptional credentials before it would be admitted amongst meteorites."*

During the period MacGregor was governor of Queensland (1909-1914), a man named Philip Bell located a large metallic stone, partly buried in the ground, on Saibai (Bell 1930:27). The section of stone above the ground measured 1.8 x 1.5 m wide and 0.9 m high and was estimated to weigh more than 50 tonnes. Bell's attempts to chip off bits of the stone or drill in it were unsuccessful. He showed the stone to MacGregor, who identified it as a meteorite. There is no further mention or discussion of the alleged meteorite, but it is significantly larger that the stone described by Hamlyn-Harris. Bell does not provide a location on Saibai for the alleged meteorite, nor any information regarding a subsequent analysis. There are no catalogued meteorites or impact craters from the Torres Strait (Meteoritical Society 2013). MacGregor visited Saibai Island on 29 April 1911 for two days as part of a tour of the Torres Strait (Anonymous 1911). It may have been during this visit that Bell led him to the stone. If this is the case, it limits the area in which the stone is located, as MacGregor did not venture far from the village of Saibai on the northwest part of the island.

Bell did not cite any oral traditions or Islander views of the stone. If the stone described here is a meteorite, could it be the source of the legend described in the previous account and not the stone analysed by Henderson? Currently, we do not know. It is likely that the stone, like the one analysed by Henderson, is simply one of many terrestrial features that are attributed to cosmic origins in the local lore.

It should be noted that stone worship was evident on Saibai and it is possible that Saibai Islanders would have revered an unusual-looking stone like the one described by Bell. For example, a sacred stone called *Adhibuya* (meaning "great light") formed the basis of a secretive warrior cult (Davis 2004:40-41). The Adhibuya stone, originally from nearby Kiwai Island in Papua New Guinea, was stolen by Saibai men for its magical powers. According to legend, the stone's mother was a virgin and its father was the moon (Haddon 1907:23). It was said to glow brightly, imparting magical powers to the warriors. Christian missionaries removed Adhibuya from the island but the narrative of the stone is still represented in a dance performance (Davis 2004:40-41). Unfortunately, the literature does not clarify where the stone was taken or where it is today.

### *Lilydale, Victoria*

An account of a cosmic impact event in Victoria is found in the traditions of the Wurundjeri people of the Melbourne region. According to Smyth (1878:456) a deep cavern at Cave Hill in Lilydale, 35 km east of Melbourne, is described in local oral traditions as a place where a star fell from the sky (Figure 2). The cavern is called *Bukkertillibe* in the Wurundjeri language, which roughly translates to "bottomless pit." According to the oral tradition, it was formed when the sky-deity *Pundjel* (more commonly known as Bunjil) became angry when the people did things that displeased him. In a rage, he caused a star to fall from the sky and strike the earth, creating the hole and killing many people. The story served to explain the origin of the cavern, which was unique in the region. The oral tradition served as a strict warning to follow laws and traditions.

Previous research by Hamacher and Norris (2009, 2010) suggests that stories of angry deities causing catastrophic events served to help maintain social order while explaining natural events. The *Bukkertillibe* oral tradition highlights a widespread theme of celestial beings throwing fiery stars to Earth as punishment for breaking sacred law. It also relates to astronomical traditions among Aboriginal groups in Victoria.

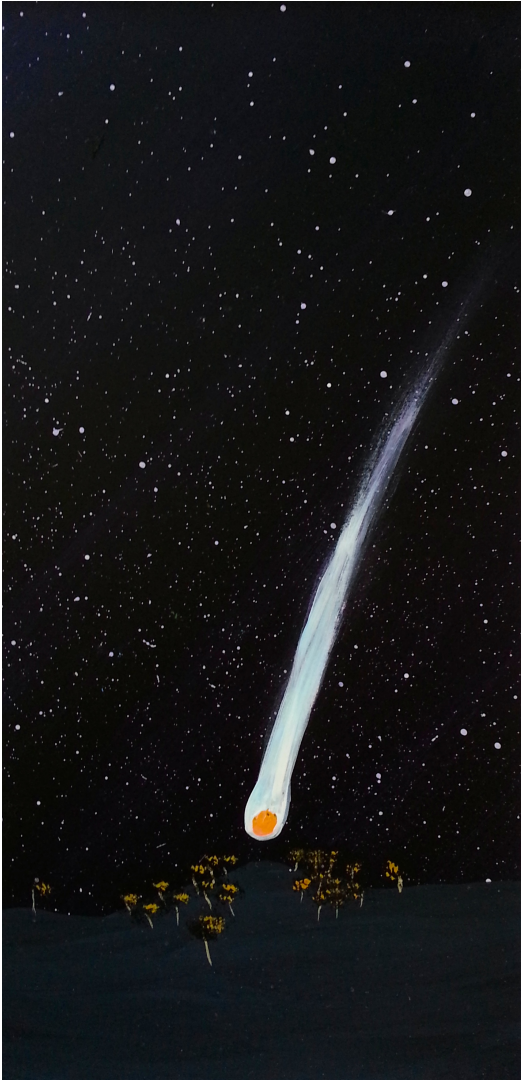

*Figure 2: "Bunjil's anger cave, Lilydale" (2000) by Tiriki Onus. Centre for Australian Indigenous Studies, Monash University, Melbourne.*

Celestial beings throwing fiery stars to Earth as punishment for breaking sacred law is found across Australia. Four examples of this include a fiery star was cast to the Earth as punishment near Lake Macquarie on the NSW Central Coast (Threlkeld 1834:51, which will be discussed later); a fire spirit in Yolngu traditions of Arnhem Land came down from the sky as a star and brought fire to the people (accidentally killing many of them) (Allen 1975:109); the Wardaman spirit *Utdjungon* (Northern Territory) that will come to Earth as a falling star to destroy the people if laws are not followed (Harney and Elkin 1949:29–31); and Ngalia men who claimed that the *Walanari* (celestial beings) cast glowing stones from the sky onto the their camp as punishment for revealing sacred knowledge to anthropologist Charles Mountford (1976:457).

Bunjil is prominent in the astronomical traditions of Aboriginal groups across Victoria. He is commonly seen as the celestial "All Father", an ancestral being responsible for teaching the people about art, life, and society (Eliade 1966). Bunjil can be seen in the sky as the star Altair (*Alpha Aquilae*) in Kulin traditions (Massola 1968:110) or the star Fomalhaut (*Alpha Piscis Austrini*) in Woiworung traditions (Howitt 1884:452).

Unfortunately, the Bukkertillibe site was destroyed when the area was excavated for a quarry (Figure 3). A sculpture commemorating the site is housed at the Lilydale campus of Swinburne University of Technology, which is adjacent to the quarry (Anonymous 2002). The sculpture, created by New Zealand sculptor Chris Booth, includes a large rock from the quarry symbolizing Bunjil. It helps to keep the story of Bukkertillibe strong despite the site's destruction.

There is no physical evidence that Bukkertillibe was the site of a cosmic impact and it is more likely that the tradition was meant to be symbolic – describing the origin of an enigmatic natural feature that served to reinforce laws and traditions. No registered meteorites are identified from Lilydale or the region (Meteoritical Society 2013), aside from the well-documented Cranbourne meteorite field in southeastern suburbs of Melbourne, 25-30 km from Lilydale (Cappadonna 2000).

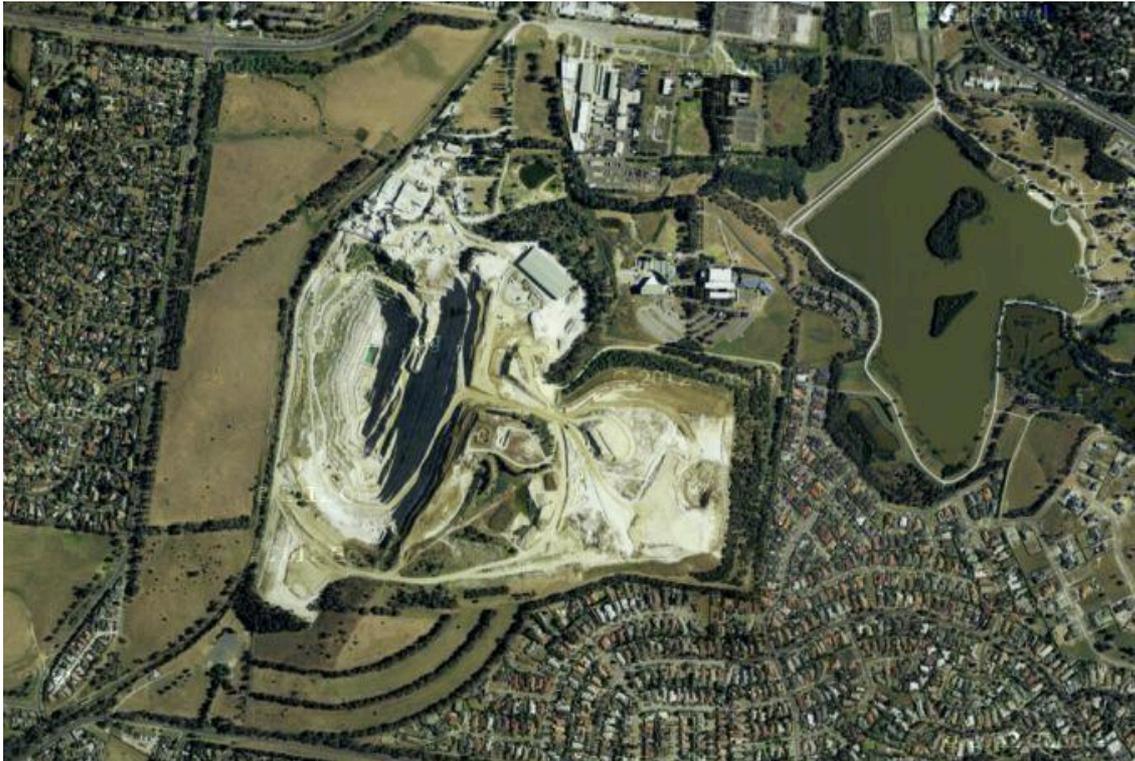

*Figure 3:* The site of Bukkertillibe, since destroyed by a rock quarry in Lilydale, Victoria. The Lilydale campus of Swinburne University is to the northeast of the quarry. Image: Google Maps.

### Helidon Springs, Queensland

A natural mineral spring, east of Toowoomba in southeast Queensland, is described in Aboriginal oral tradition as the place where a hunting party was camped when a large stone fell from the sky, killing many people. It is where the star fell that the mineral spring burst forth (Anonymous 1881). Another oral tradition of the spring (Meston 1899) attributes the death of the people to a flood instead of a falling star:

*"The Helidon district they called Yabarba, the name of the Curriejung* [tree], *and the spring was known as Woonarrajimmi, the place 'where the clouds fell down'. In a former age a numerous crowd of blacks were camped on the present spring, and a gin was standing by the fire scratching her head, from which she extracted two of the species Pediculus capitis* [head lice]. *While surveying these captives in the palm of her hand, a puff of wind blew them into the fire, an unhappy accident, always attended by penalties terrible to contemplate. Celestial vengeance on this awful occasion was satisfied only by the clouds falling and burying the whole tribe fathoms deep in the earth. From the buried tribe sprang the*

*Helidon spring, the waters of which they call 'kowoor,' regarded as a highly efficacious bath for sick blacks, but not to be used as a beverage under any possible circumstances, the reason being clearly and logically dellned."*

The area is prone to flooding, which may form part of the basis of the tradition. It is possible that the description of a falling star was a misinterpretation of the story by the non-Indigenous person who recounted the story. It is also plausible that Anonymous (1881) was simply told one version of the story. The tradition is similar to one from the Awabakal people on the Central Coast of New South Wales, who have traditions about an event that occurred at Lake Macquarie (called *Kurra Kurran*) (Threlkeld 1834:51, Hamacher and Norris 2009:69). In the tradition, people killed lice by roasting them on a fire. This angered a sky being, which took the form of a giant goanna. He cast down a fiery stone from the sky that killed many people at Fennel Bay. Bits of petrified wood that jet out of the soil represent fragments of the preserved fallen stone.

Another detailed account of this oral tradition is provided in Anonymous (1913), but does not mention the fall of a stone or the burning of lice. Instead, it describes the flood originating from a flood. According to this version, the Aboriginal people of Brisbane called Helidon Spring *Gooneol Goong* meaning "water from the moon". The Aboriginal people near what its now Toowoomba called the spring *Woourrajimigh,* meaning "the place where the clouds fall down." The identity of the person who collected the story and the Aboriginal person(s) who shared it are not given, so the account must be taken cautiously. The close association of the first oral tradition with similar Aboriginal traditions along the eastern coast indicates that the oral tradition is not a fabrication, although there seem to be multiple variations of the tradition. The site is now a caravan park and there are no recorded meteorite finds from the area.

*Other Accounts*

Lake Argyle in the far northeastern corner of Western Australia was formed when the Australian government dammed the Ord River in 1963. In 1986, Aboriginal artist Rover Thomas (c1926-1998) said the place that is now Lake Argyle was where a star fell to Earth long ago (Deutscher and Hackett n.d.). No meteorites or impact craters are registered from the area.

In some cases, Aboriginal communities attributed cosmic origins to terrestrial objects. In 1885, a large half-ton coral "stone" was identified by a constable north of Carnarvon, Western Australia that local Aboriginal people claimed "fell from the moon" (Anonymous 1885a,b).

**DISCUSSION**

Across Australia, transient celestial phenomena, such as meteors, comets, eclipses, and aurorae, were generally seen as omens of death and disease or attributed to the actions of spirits and evil beings (Hamacher and Norris 2010, 2011a,b, Hamacher 2013, respectively). This explains the generally negative and fearful reaction of Indigenous people to witnessing significant meteoritic events such as airbursts (exploding meteors) and meteorite impacts. Examples not described in previous research include airbursts over Bairnsdale, Victoria in May 1880 (Melbourne Chamber of Commerce 1880:3), Currawillinghi homestead in far southern Queensland in September 1890 (Anonymous 1890), and Bellenger Heads in northern coastal New South Wales in June 1899 (Anonymous 1899). Near misses - where people were nearly struck by falling meteorites - are recorded from the Herbert River Valley in northern Queensland in November 1882 (Lumholtz 1889:175-176) and the Swan River in Perth, Western Australia in July 1838 (Anonymous 1838:3). These are only a few examples of many recorded in historical documents.

It should be emphasized that many Indigenous Australians, particularly those in more remote regions, maintain strong traditions despite 225 years of colonisation, Christian conversion, and Western education. Sometimes these traditions mix with Western understandings of the world and highlight issues with cultural sensitivity, even among Aboriginal people themselves.

While working as a consultant in Kakadu National Park (Northern Territory) the author and a colleague developed an Aboriginal astronomy night tour for a local Indigenous-owned tourism company. While training the tour guides (most of whom were Aboriginal) during a night session, two bright meteors streaked across the sky. One of the Aboriginal guides said that in local Gagudju traditions, the appearance of a meteor signified that someone had died. Although the tour guides were lighthearted about the appearance of the meteors, the following morning met with somber news: during the night, two members of the local Aboriginal community had passed away. Some of the training had to be postponed while the Aboriginal guides were engaged in "sorry business" (a period of ceremony and mourning when someone passes away).

There is little doubt that the guides discussed the appearance of the two meteors and it is possible that the event reinforced a cultural perception that meteors were portents of death. This event highlights the importance of acknowledging cultural sensitivity and serves to show the influence of oral traditions on contemporary Aboriginal cultures.

**SUMMARY**

This report analyzes 11 oral traditions and historical records (in eight groups) that have been identified since the Hamacher and Norris (2009) study, which challenges the previous conclusion that no meteoritic events recorded in oral tradition led to the (re)discovery of meteorites and/or craters that were previously unknown to Western science.

The oral traditions of the Narraburra, Henbury, Monte Collina, and Murnpeowie meteorites are plausible instances of Aboriginal oral traditions leading to (or coinciding with) colonial identification of the meteorites (supporting hypothesis #2). H.C. Russell confirms this in regards to the Narraburra meteorite, but the Murnpeowie and Monte Collina meteorites are unconfirmed.

Aboriginal views of Henbury are only recorded in the literature after it was suspected of being an impact site, but the presence of oral traditions describing its formation suggest its origin was known to Aboriginal people before colonisation (supporting hypotheses #1 and #2). The Henbury account is important in demonstrating the longevity of oral traditions. If the tradition is a living memory of the event, it is well over 4,500 years old (supporting hypothesis #3). Unlike Henbury, analysis of the Narraburra meteorite's exposure to Earth's atmosphere suggests the fall was not witnessed, but rather inferred.

The accounts from Saibai, Lilydale, Benyeo, Monte Barcoola, Helidon Springs, Lake Argyle, and Carnarvon are not associated with a confirmed meteorite or impact crater. Some of these accounts could potentially be confirmed (such as the alleged meteoritic materials used to build the Benyeo Homestead, the (re)discovery of a meteorite on Saibai, or a meteorite west of Monte Barcoola.

The descriptions of Lilydale, Lake

Argyle, Carnarvon, and Helidon Springs are not likely to be associated with a meteorite or crater. These seem to be either misidentifications or mythological/symbolic in nature.

This is not conclusive and future research could shed more light on the relationship between these places and oral traditions of meteoritic events.

## ACKNOWLEDGEMENTS

I would like to acknowledge all Aboriginal and Torres Strait Islander elders and custodians, and thank Steve Hutcheon, John Carlson, Tui Britton, and the Monash University Indigenous Centre. This research was funded by the Australian Research Council project DE140101600.

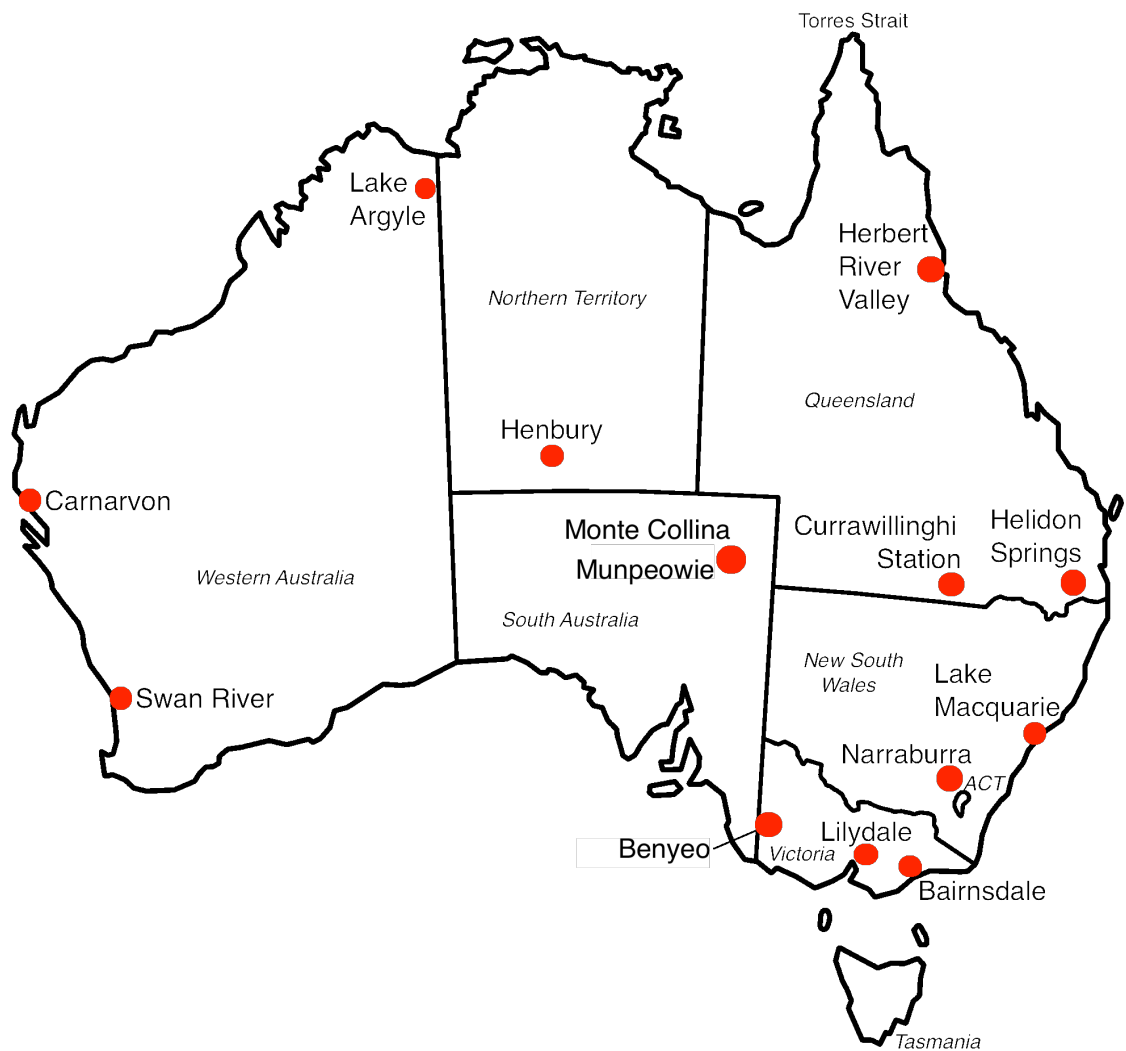

*Figure 3: Places described in the text. Map by the author.*

## REFERENCES

Alderman, Arthur R.
1932    The meteorite craters at Henbury, Central Australia. *Mineralogical Magazine* 23: 19-32.

Allen, Louis A.
1975    *Time before Morning: Art and Myth of the Australian Aborigines*. Thomas Y. Crowell, New York.

Anonymous
1838   Meteors. *The Colonist* (newspaper, Sydney, NSW: 1835-1840), Wednesday, 11 July 1838: 3.
1881   City Police Court. *The Queenslander* (newspaper, Brisbane, QLD: 1866-1939). Saturday, 1 January 1881: 25.
1885a  Discovery of a meteorite at Carnarvon. *The Daily News* (newspaper, Perth, WA: 1882-1950), Friday, 8 May 1885: 3.
1885b  *The West Australian* (newspaper, Perth, WA: 1879-1954), Thursday, 21 May 1885: 3.
1890   The Strike. *Western Star and Roma Advertiser* (newspaper, Toowoomba, QLD: 1875-1948), Wednesday, 3 September 1890: 2.
1899   A Large Meteor. *The Clarence River Advocate* (newspaper, Maclean, NSW: 1898-1949), Friday, 17 March 1899: 3.
1910   The Murnpeowie meteorite: a world wonder. *The Register* (newspaper, Adelaide, SA: 1901-1929). Wednesday, 6 April 1910: 8.
1911   The Governor's Tour: Visit to the North. *The Brisbane Courier* (newspaper, Brisbane, QLD: 1864-1933), Tuesday 23 May 1911: 5.
1913   Waters of the moon. *The Western Champion and General Advertiser for the Central-Western Districts* (newspaper, Barcaldine, QLD: 1892-1922). Saturday, 13 December 1913: 3.
1934   To investigate meteorite craters. *The Advertiser* (newspaper, Adelaide, SA: 1931-1954), Thursday, 4 January 1934: 9.
2002   [*The Sculpture. Bukker Tillibul – What's in a Name?*](#) Swinburne University of Technology, Lilydale Campus.

Barham, Anthony J., Michael J. Rowland, and Garrick Hitchcock
2004   Torres Strait bepotaim - an overview of archaeological and ethnoarchaeological investigations and research. *Memoirs of the Queensland Museum* 3(1): 1-72.

Bell, Philip J.
1930   Huge Meteorite. *Sunday Mail* (newspaper, Brisbane, QLD: published from 1926-1954), 1 June 1930: 27.

Bevan, Alex W.R.
1992   Australian meteorites. *Records of the Australian Museum (Supplement)* 15: 1-27.

Bevan, Alex W. R., and Peter Bindon
1996   Australian Aborigines and Meteorites. *Records of the Western Australian Museum* 18: 93–101.

Buchwald, Vagn F.
1975   *Handbook of Iron Meteorites, Vol. 3*. University of California Press, pp. 875-877.

Cappadonna, William J.
2000   *A History of the Cranbourne Meteorites*. City of Casey, Victoria.

Chang, C. and H. Wänke
1969   *Beryllium-10 in Iron Meteorites, Their Cosmic-Ray Exposure and Terrestrial Ages.* In "*Meteorite Research*," edited by P.M. Millman. Astrophysics and Space Science Library Volume 12: 397-406. Riedel Publishers, Dordrecht.

Davis, Richard
2004   *Woven History Dancing Lives: Torres Strait Islander Identity, Culture And History*. Aboriginal Studies Press, Canberra.

Deutscher and Hackett
n.d.   [*24 Rover Thomas*](#) *(Joolama) (c1926—1998): Lake Argyle, 1994*.

Dunai, Tibor J.
2010   *Cosmogenic Nuclides: Principles, Concepts and Applications in the Earth Surface Sciences.* Cambridge University Press.

Eliade, Mercia
1966   Australian Religions, Part 1: An introduction. *History of Religions* 6(2): 108-268.

Elkin, Adolphus Peter
1949   *Aboriginal Men of High Degree*. University of Queensland Press, Brisbane.

Gale, J.
1924   Then and Now – Young and Its Surrounding Country: As it was in A.D. 1857-59 – As It Is A.D. 1924. *Daily Witness* (newspaper, Young, NSW), Monday, 21 January 1924: 4.

Gibbons, G.S.
1977   Index of Australian meteorites. *Australian Journal of Earth Sciences* 24(5): 263-268.

Gill, T.
1926 Unrecovered meteorite. *The Mail* (newspaper, Adelaide, SA: 1912-1954). Saturday, 10 April 1926: 17.

Goldsmith, John M.
2000 Cosmic Impacts in the Kimberly. *Landscope Magazine* 15(3): 28–34.

Grady, Monica M.
2000 *Catalogue of Meteorites*. Cambridge University Press, London.

Grassie, James
1888 Stray Notes: Meteoric Stones. *Border Watch* (newspaper, Mount Gambier, SA: 1861-1954), Saturday, 30 June 1888: 4.
1898 A meteorite. *Border Watch* (newspaper, Mount Gambier, SA: 1861-1954), Saturday, 22 January 1888: 3.

Haddon, Alfred C. (editor)
1904 *Reports of the Cambridge Anthropological Expedition to Torres Straits, Vol. 5: Sociology, Magic and Religion of the Western Islanders*. Cambridge University Press, Cambridge.

Hamacher, Duane W.
2013 Aurorae in Australian Aboriginal traditions. *Journal of Astronomical History & Heritage* 16(2): 207-219.
2011 Meteoritics and Cosmology among the Aboriginal Cultures of Central Australia. *Journal of Cosmology* 13: 3743–3753.

Hamacher, Duane W. and John M. Goldsmith
2013 Aboriginal oral traditions of Australian impact craters. *Journal of Astronomical History and Heritage* 16(3): 295-311.

Hamacher, Duane W. and Ray P. Norris
2009 Australian Aboriginal Geomythology: eyewitness accounts of cosmic impacts? *Archaeoastronomy* 22: 60–93.
2010 Meteors in Australian Aboriginal Dreamings. *WGN - Journal of the International Meteor Organization* 38(3): 87–98.
2011a Comets in Australian Aboriginal astronomy. *Journal of Astronomical History and Heritage* 14(1): 31-40.
2011b Eclipses in Australian Aboriginal astronomy. *Journal of Astronomical History and Heritage* 14(2): 103-114.

Hamlyn-Harris, R.
1913 Ethnographic notes of the Torres Strait. *Memoirs of the Queensland Museum, Vol. 2*. Anthony James Cumming, Brisbane.

Harney, William Edward and Adolphus Peter Elkin
1949 *Songs of the Songmen: Aboriginal Myths Retold*. F. W. Cheshire, Melbourne.

Heymann, Dieter
1965 Rare gas evidence for two paired meteorite falls. *Geochimica et Cosmochimica Acta*, 29(12): 1203-1208.

Hill, Ernestine
1937 Stardust for sale, money in meteorites: bombardment of Australia. *The Sydney Morning Herald* (Sydney, NSW: 1842-1954), Saturday, 4 December 1937: 7.

H.J.L.
1926 About meteorites. *The Mail* (newspaper, Adelaide, SA: 1912-1954), Saturday, 8 May 1926: 17.

Hodge-Smith, T.
1939 Australian meteorites. *Australian Museum Memoir* 7: 1–84.

Howitt, Alfred William
1884 On Some Australian Ceremonies of Initiation. *The Journal of the Anthropological Institute of Great Britain and Ireland* 13: 432–459.

Liversidge, A.
1903 The Narraburra Meteorite. *Journal and Proceedings of the Royal Society of New South Wales* 37: 234-242.

Lumholtz, Carl
1889 *Among Cannibals: an account of four years' travels in Australia and of camp life with the Aborigines of Queensland*. John Murray, London.

Mason, Brian
1974 Notes on Australian meteorites. *Records of the Australian Museum* 29(5): 169–186.

Massola, Aldo
1968 *Bunjil's Cave: Myths, Legends and Superstitions of the Aborigines of South-East Australia*. Lansdowne Press, Melbourne.

Melbourne Chamber of Commerce
1880    Melbourne Chamber of Commerce Committee's Annual Report. *Camperdown Chronicle* (newspaper, Camperdown, VIC: 1877-1954), Tuesday, 4 May 1880: 3.

Meston, A.
1899    Heldion. *The Queenslander* (newspaper, Brisbane, QLD: 1866-1939). Saturday, 11 February 1899: 18.

Meteoritical Society
2013    Meteoritical Bulletin Online Database. Lunar & Planetary Institute, Houston, TX.

Mountford, Charles Pearcy
1976    *Nomads of the Australian Desert*. Rigby, Adelaide.

Russell, Henry Chamberlain
1890    Description of the Narraburra meteor. *Journal and Proceedings of the Royal Society of New South Wales* 24: 81-82.

Smith, L.L.
1910    An Australian meteorite. *American Journal of Science* 30(178): 264-266.

Spencer, L.J.
1934    Murnpeowie (South Australia), a granular type of meteoric iron. *Mineralogical Magazine* 24: 13-20.

Smyth, R. Brough
1878    *The Aborigines of Victoria, Volume 1*. J. Ferres, Melbourne.

Thorpe, W.W. and Fredrick D. McCarthy
1958    Aboriginal names: their native meanings. *Dawn: Journal for the Aboriginal People of New South Wales* 7(6): 13-19.

Threlkeld, Lancelot Edward
1834    *An Australian Grammar: Comprehending the Principles and Natural Rules of the Language as Spoken by the Aborigines in the Vicinity of Hunter's River, Lake Macquarie, New South Wales*. Stephens & Stokes, Sydney.

Victorian Heritage Database
1974    Benyeo Homestead. Property No. B2847. Victorian Heritage Database.

Vitaliano, Dorothy B.
1973    *Legends of the Earth*. Indiana University Press, Bloomington.

Vox
1945    Out among the people: from the inland. *Chronicle* (newspaper, Adelaide, SA), Thursday, 25 January 1945: 35.

**AUTHOR BIOGRAPHY**

Dr Duane Hamacher is an academic and ARC Discovery Research Fellow in the Nura Gili Indigenous Programs Unit at the University of New South Wales in Sydney, Australia. His teaching and research focuses on cultural and historical astronomy, with an emphasis on Indigenous Australia and Oceania.